\documentclass[12pt]{article}
\usepackage{mod_moriond}
\usepackage{float}
\usepackage{equationarray}
\usepackage{tabularx}
\usepackage{supertabular}
\usepackage{epsfig}
\usepackage{psfig}
\usepackage{wrapfig}
\def\simlt{\lower.5ex\hbox{$\; \buildrel < \over \sim \;$}}
\def\simgt{\lower.5ex\hbox{$\; \buildrel > \over \sim \;$}}
\begin{document}
%
%\initfloatingfigs
%
%
\heading{GOHSS: a Fibre--fed Multiobject NIR Spectrograph\\ for
the Italian Galileo Telescope}
\author{R.~Scaramella$^{1}$, E.~Cascone$^{2}$, F.Cortecchia$^{2}$,
R.S.~Ellis$^{3}$,  D.Lorenzetti$^{1}$,}{D.Mancini$^{2}$,
I.~Parry$^{3}$, F.~Pich\'e$^{3}$, F.~Vitali$^{1}$}
{$^{1}$ Osservatorio Astronomico di Roma, Monteporzio Catone, Italy.}
{$^{2}$ Osservatorio Astronomico di Capodimonte, Napoli, Italy.}
{$^{3}$ Institute of Astronomy, Cambridge, England.}
\begin{moriondabstract}
We present the concept and preliminary design of the Galileo OH
Subtracted Spectrograph [GOHSS], a multifibre NIR spectrograph for faint
objects. The instrument represents a collaboration between the
Institute of Astronomy, Cambridge [IoA] and the Observatories of
Naples and Rome and will be a second--light instrument for the 3.6m 
Galileo telescope [TNG] located on La Palma.

The NIR spectrograph accomplishes OH night-sky suppression in a different
way from the hardware solution used by both OHS \cite{Maiharaetal} and 
COHSI \cite{COHSI}. GOHSS provides a multiechelle design with software 
subtraction capable of yielding $\sim 28$ spectra in J+H bands at a 
spectral resolution ${\cal R} \sim 3000$.  Such a resolution is the
minimum necessary to reduce the impact of atmospheric OH lines.

\end{moriondabstract}

\section{Scientific Aims of the Instrument}
A number of key scientific programs can benefit from a significant
reduction in the background in the J and H bands. These can be essentially 
grouped into three classes: {\sl redshifts and spectral studies of galaxies, 
spectral properties of quasars, the study of faint, cold or obscured
stars}. In the following we will illustrate some programmes in the first of
these categories.\\

\noindent
\emph{High Redshift Galaxies, Their Counts, Properties and Evolution}\par
\medskip
In recent years there has been considerable progress in determining the
counts, colours and redshift distributions of faint galaxies. Such studies 
are crucial for the theory of structure formation. The current limits for 
ground-based counts are $B \sim 27 \div 28~mag$ \cite{Metcalfeetal}, while
the 4-m redshift samples have painstakingly reached $B \sim 24~mag$ 
\cite{Ellisetal}, $I \sim 22~mag$ \cite{CFRS} which represents a hard 
instrumental limit in the visible band. Further progress has been possible
through the use of the \emph{unique combination} of \emph{Keck + HST}.  HST 
has provided new insight into the morphological appearance of high $z$ 
galaxies including those faint sources in the publically-available HDF. Some
high $z$ candidates located with Lyman-limit imaging techniques 
\cite{SteidelHamilton} have been spectroscopically confirmed by Keck 
exposures \cite{Steideletal}. A gap emerges, however, between the progress
to $z\simeq$1 secured from 4-m surveys and limited data now available
at $z>$2. From various considerations, the bulk of the star formation
activity appears to have occurred in this little-explored region for
which NIR spectroscopy is crucial.
\par
\smallskip
\noindent
The main scientific targets of GOHSS are:
\begin{itemize} 
\item[1.] \emph{Redshift Measurements of Faint Galaxies}. For $z > 1.1$
the [OII]3727 emission line (a principal line for the redshift determination) 
leaves the visible band, whereas to observe the Ly${\alpha}$ (often very 
weak or absent) requires $z > 2.2$. Consequently \emph{without low 
resolution spectroscopy in the J and H bands, it is very hard to determine the 
redshifts of normal galaxies in the range $1 \leq z \leq 2$}. A multifibre 
device would make practical the simultaneous spectroscopy in the same exposure.
\item[2.] \emph{Study of the Star Formation Rate (SFR) at High Redshift}.
Spectroscopy in the J and H bands allows the measurement of the H${\beta}$
line (providing the ionization level) and the equivalent width (EW) of the 
H$\alpha$ line in the range $0.5 \le z \le 2$. The latter line is {\sl one 
of the most reliable tracers of star formation} and its measurement in
many distant galaxies would provide crucial information in understanding
the rates, timescales and modes of galaxy formation including environmental 
effects, the role of different IMFs etc. 
\item[3.]\emph{Redshift Measurements of Gravitational Arcs}. Over the past
few years, spectacular examples of gravitational lensing have been provided 
through the identification of giant arcs. Such arcs arise from spacetime 
curvature induced by the gravitational field of the dark matter in rich 
galaxy clusters. Their importance is twofold. Firstly, when the redshift
of the lensed object is known, it is possible to accurately model the
potential well of the cluster and determine the dark matter distribution.
Secondly, the cluster itself acts as a {\sl natural} telescope: the lensing 
phenomenon amplifies the apparent luminosity of lensed objects which become 
observable at greater distances than unlensed {\sl field} objects. However,
spectroscopy of such arcs present a challenge to most spectrographs as the 
arc morphologies are not easily accommodated within a single long-slit.
This problem can be alleviated by using an Integral Field Unit [IFU].
\item[4.]\emph{Galaxy Redshift Measurements along the Line of Sight to
Quasars}. An efficient way to find galaxies at high redshift is to
perform imaging observation in the K band around quasars at medium
redshift, or to select objects in absorption with metallic lines
(typically CIV) along the line of sight. In this case, the infrared
spectra of these objects confirm which source is the absorber as well
as yielding useful information about the SFR. In this case, an IFU used
so that the spectral width $\Delta \lambda$ is centred on important lines 
in the damped absorbing cloud restframe could yield interesting results.
\end{itemize}

\section{Technical Solutions}

\subsection{Hardware {\sl vs} Software Solution}

In undertaking a feasibility study of the instrument, we have quantitatively
compared the hardware \emph{suppression vs} software \emph{subtraction}.  The
former solution involves the suppression of OH lines by means of a physical
mask at an intermediate high resolution spectrum and the subsequent
recombination in {\sl white light} to feed a low resolution
imager-spectrograph. Such a solution has been implemented by the Japanese
group at Kyoto \cite{Maiharaetal}~\cite{Iwamuroetal} and by Io in the case of
COHSI \cite{COHSI}.

The second solution adopts a high resolution ($ {\cal R} \simgt 3000$) 
approach which allows the registration of the spectrum directly on the 
detector. By selecting regions uncontaminated by the OH lines and 
{\sl rebinning} in case of weak signals, the need for hardware suppression
is avoided. This solution, which offers significant advantages, has not 
been implemented thus far because of the high costs of the large format IR 
detectors and their relatively high dark current. However, the situation 
with regard to these problems is rapidly changing and this progress makes
the {\sl software} suppression method now preferable. Other, less direct
methods e.g. blocking filters do not appear yet satisfactory \cite{Herbst}, 
\cite{Jonesetal}. We note that the software solution is also envisaged for the 
VLT \cite{VIRMOS}, although with the use of slits instead of fibres, a much 
larger multiplexing than adopted here and, of course, at considerably increased
cost.
%
%%%%%%%%%%%%%%%%%%%%%%%%%%%%%%%%%%%%%%%%%%%%%%%%%%
\begin{figure}[h]
\centerline{ %
\epsfig{figure=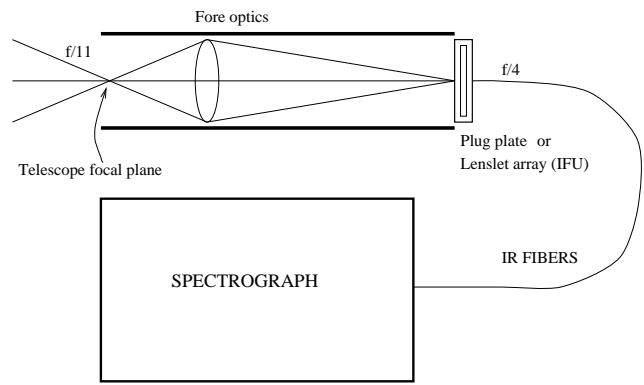,width=5truecm,angle=270}
}
\caption{ \label{fig:schema}  \footnotesize  \baselineskip 0.4cm %
	Main modules of GOHSS }
\end{figure}
%%%%%%%%%%%%%%%%%%%%%%%%%%%%%%%%%%%%%%%%%%%%%%%%%%

\subsection{Optical Design}

The optical design of GOHSS has passed through several reviews and
modifications since its first conceptual design based on that of COHSI, 
the (hardware) OH Suppressor Spectrograph nearing completion at IoA. The 
scientific aims and conceptual requirements described above lead to a 
modular design. In Figure~\ref{fig:schema} we show the overall layout of 
the three main components: the fore--optics, the IR fibres and the 
dispersing optics.

The fore--optics section is intended to modify the f/11 beam of the 3.5m
Galileo telescope and provide a suitably enlarged scale to allow the 
positioning of the IR fibres along with their interface optics, {\sl i.e.}  
the lenslets. The fore--optics consists of two simple doublets. By replacing 
the first doublet and readjusting the lens separation, the spatial scale can 
be varied. This is likely to be extremely useful in transporting the 
instrument to different telescopes, such as the LBT.

% DA CIRPASS, tranne ECHELLOGRAM su 3 pixels
%
%
%%%%%%%%%%%%%%%%%%%%%%%%%%%%%%%%%%%%%%%%%%%%%%%%%%
\begin{figure}[h]
\centerline{ %
\psfig{figure=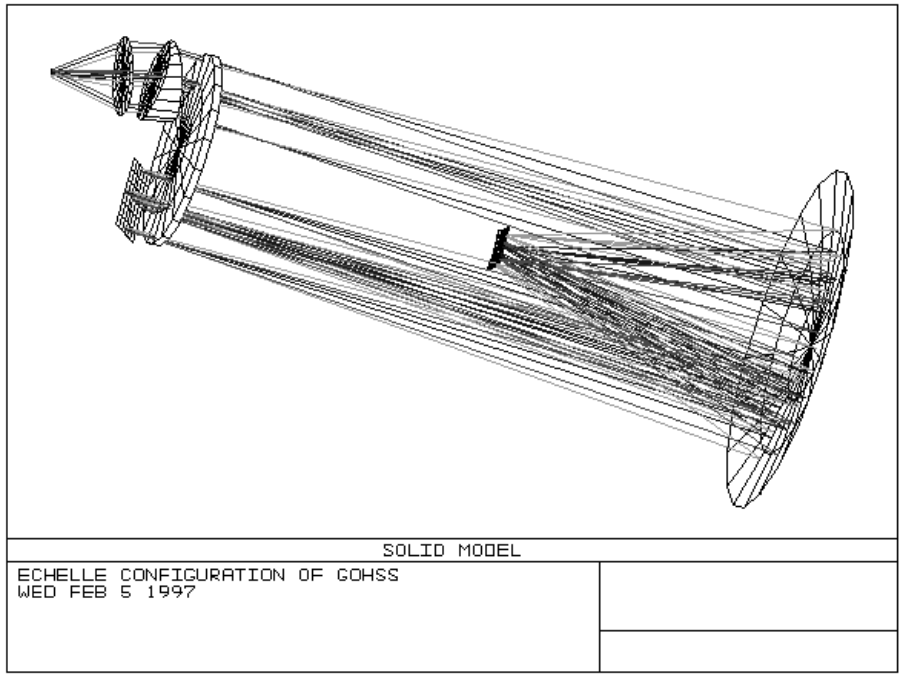,width=12truecm}
}
\caption{ \label{fig:sideview}  \footnotesize  \baselineskip 0.4cm %
	GOHSS spectrograph optical layout}
\end{figure}
%%%%%%%%%%%%%%%%%%%%%%%%%%%%%%%%%%%%%%%%%%%%%%%%%%

In multiobject mode GOHSS will most likely have a plug-plate fibre system 
with a single fibre per galaxy. To feed the light into the fibre efficiently 
a micro--lens is used to image the telescope pupil on to the fibre at f/4,
significantly reducing focal ratio degradation (FRD).

%The advantages offered by this kind of matching with respect to directly
%re--image of the telescope focus are discussed by Hill et al. (1983)
%and Haynes and Parry (1994). 

A small field stop ahead of the microlens and in the telescope focal
plane samples $\sim \, 1.5''$. With a final camera focal ratio of f/1.2, 
a single fibre corresponds to two pixels. Allowing for some residual FRD
and possible misalignment in the optics, conservatively we will adopt three 
pixels for the spatial coverage.

The instrument specifications are defined primarily by the scientific aims.
For the optical design, the most important specifications are:

\begin{itemize} 
\item Spectral resolution of $\sim$ 3000.  This is dictated by the nature 
of the OH sky spectrum. This criteria could be lessened somewhat when 
observing in the J band where the OH lines are more sparse. In this case,
it is possible to trade spectral resolution for extension towards shorter
wavelengths. 
\item The MOS mode requires a fibre aperture of $\sim 1.5"$ because of 
the inherent size of the target galaxies. A minimum of $20 - 25$ fibres
is necessary to provide significant gains.
%
%\item The wavelength-coverage $\times$ multiplex-gain product 
%[(pixel/fiber) $\times$ (number of fibers) $\times$ (spectral band width)]
%which is essentially a measure of the information content ({\sl i.e.} the
%extent of the equivalent spectral band covered by all the available
%pixels);
%
\item The simultaneous spectral coverage should be 1 $\div$ 1.8 $\mu m$.
\end{itemize}

%
%%%%%%%%%%%%%%%%%%%%%%%%%%%%%%%%%%%%%%%%%%%%%%%%%%
\begin{figure}[h]
\centerline{ %
\epsfig{figure=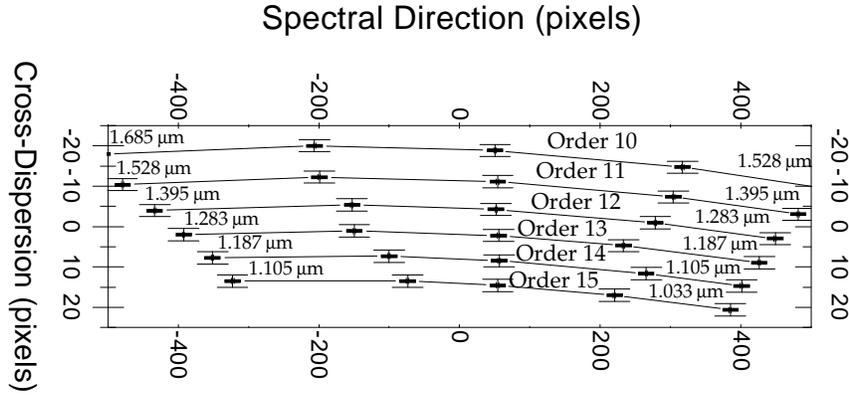,width=6truecm,angle=270,clip=}
}
\caption{ \label{fig:sixorders}   \footnotesize  \baselineskip 0.4cm %
Details of the order layout for the fibre/object situated at the
centre of the ``slit''. Error bars show the expected extent of the
data, both spectrally and spatially (three $\times$ three pixels) 
}
\end{figure}
%%%%%%%%%%%%%%%%%%%%%%%%%%%%%%%%%%%%%%%%%%%%%%%%%%

%%%%%%%%%%%%%%%%%%%%%%%%%%%%%%%%%%%%%%%%%%%%%%%%%%
%We have comparatively analyzed different ways to disperse the J and H
%band onto one, two or four arrays, both bands simultaneously or
%separately: this has produced three different designs of the
%spectrograph: the Baranne configuration, the Classical configuration,
%and a cross-dispersed echelle. The conclusion is that {\bf the most
%promising configuration is the cross-dispersed echelle}, which will be
%described in the following.
%
%\begin{wrapfigure}{r}{6truecm}
%\mbox{\epsfig{figure=dum.ps,width=5truecm}} %
%%
%\caption{\label{fig:sixorders}   \footnotesize  \baselineskip 0.4cm %
%Details of the orders layout for the fiber/object situated at the
%center of the ``slit''. The error bars show the expected extent of the
%resolution element, both spectrally and spatially (three $\times$
%three pixels)} 
%%
%\end{wrapfigure}
%%

The conclusion from several optical designs studied by I.~Parry and
F.~Pich\'e at IoA is that the most promising configuration is the
cross-dispersed echelle.

The optical scheme is shown in Figure~\ref{fig:sideview}. Fibres from the 
telescope are aligned along a slit at the focal plane of a Schmidt camera
(Figure~\ref{fig:sideview}). The f/5.5 primary mirror generates a 150 mm 
diameter beam which is then projected onto the grating. The spectrum is 
then projected onto the relay mirror (situated at the Schmidt camera focus) 
and re--collimated by the primary mirror. It then passes through a 
cross--disperser prism to properly distribute the orders onto the detector. 
After cross-dispersion the spectrum is imaged onto the detector by an 
f/1.2 camera. The detailed layout for a single fibre spectrum is shown in 
Figure~\ref{fig:sixorders}.

The number of photons per pixel detected from the sky background is
very low so it is vital that GOHSS contributes significantly less than
this in terms of the background due to thermal emission from within
the spectrograph.  From detailed considerations we find that in order
to keep the thermal background to an acceptably low level, the
temperature of the spectrograph has to be maintained between $\approx$
217K and 236K or $\approx$ -56$^o$C to -37$^o$C.

\section{Expected performances}

Assuming the following efficiencies for: Atmosphere $+$ TNG:
$\epsilon_1 \simeq 68\%$; multifibre system: $\epsilon_2 \simeq 81\%$;
spectrograph: $ \epsilon_3 \simeq 64\% $; camera: $\epsilon_4 \simeq
40\%$, we obtain total efficiencies of: $\epsilon_{Tot} =\Pi_{i=1}^4\,
\epsilon_i\ \, \approx\, 14\%$. Now, with a 4$m$ class telescope 
(TNG) the software solution allows to work in BLIP (Background Limited
Instrumental Performance) conditions in the J and H bands once the
detector {\sl dark current} [{\sl DC}] is below 0.05 $e^-/s$
\cite{Hodappetal}. With a 8$m$ class telescope (VLT), the instrument
would always be in BLIP conditions, even with high {\sl DC} values
($>$ 0.15 $e^-/s$) and low overall efficiencies ($< 10\%$). Therefore,
the relative efficiency {\it {\sl S/W} vs {\sl H/W} } is always $>$ 1
in case of low {\sl DC}, before considering possible OH scattering
problems associated with the hardware mode.

%%%%%%%%%%%%%%%%%%%%%%%%%%%%%%%%%%%%%%%%%%%%%%%%%%
\begin{figure}[h]
\centerline{ %
\psfig{figure=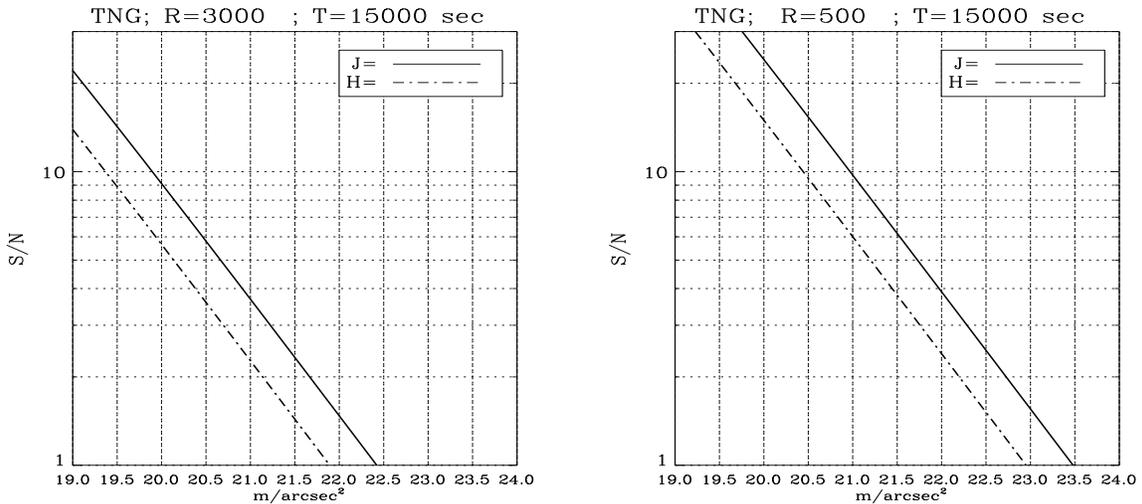, %
	width=16truecm,height=7truecm}
}
\caption{ \label{fig:TNGSNR} {  \footnotesize  \baselineskip 0.4cm %
	We show the S/N vs limiting brightness for two different
	resolutions in the TNG case. 
	The values assumed are $\epsilon_{Tot}=14\%$ and 
	dark=0.1 e/s}
	% The upper panel has dark=0.1 e/s, the lower one dark=0.01 e/sec.}
}
\end{figure}
%%%%%%%%%%%%%%%%%%%%%%%%%%%%%%%%%%%%%%%%%%%%%%%%%%
The SNR in the left panel of Fig.~\ref{fig:TNGSNR} does not takes into
account the possibility of software rebinnig (or more advanced methods
to extract the signal). Nonetheless, an increase in SNR of a factor 
$\sqrt{f_{rebin}}$ is realised, where $f_{rebin}\simeq R_c^{orig}/R_c^{low}$. 
Furthermore, it can be $f_{rebin} \sim 6$ when $R_c^{orig} \sim 3000$ and 
$R_c^{low} \sim 500$. This is shown in the right panel of 
Fig.~\ref{fig:TNGSNR}, where we show {\it the continuum SNR} as a function 
of object brightness under the assumption of fibres of area $1''$ square 
(diameter of $1.3''$). Notice that a ``typical'' faint galaxy has angular 
extent of $\simlt 2''$, therefore an approximate conversion in limiting 
magnitudes is given by subtracting $\sim 1$ mag from the abscissa scale. This 
is compensated by the fact that most of the information is typically stored 
in lines of remarkable EW, an aspect which is being studied with more
detailed simulations of spectra. 

It is important to note that, when one limits to a single band (i.e. J or H), 
\emph{the GOHSS multiplexing could double}, reaching a number of possible 
targets of $\sim 50$ in a single exposure. This can be exploited in the 
study of distant clusters of galaxies, where the main aim is to determine if 
a galaxy belong to the cluster. Moreover, it becomes feasible to study the 
\emph{SFR as a function of density and time}, because the $H_\alpha$ line 
can be studied for cluster members and compared to the general behaviour
in the field, a spectral counterpart of the Butcher--Oemler effect. The 
multifibre mode is then essential, given the crowding of targets.
%a nice example is given by the high $z\sim 0.83$ ms1054 EMSS
%cluster, for which the $H_\alpha$ is observable at 
%$\lambda \sim 1.2 \, \mu m$.
%
%The modularity of GOHSS (see Fig.~\ref{fig:schema}) also opens up the
%possibility of moving the instrument to larger telescopes, such as
%LBT, with changes in the preoptics.

% References listed in alphabetical order ...

%
\vfill
\end{document}